\documentclass[amssymb,nofootinbib,nobibnotes,aps,prd,preprintnumbers]{revtex4}

 \usepackage{here}

\usepackage{amssymb,amsfonts,amsmath,bm,color,multirow,cases,empheq,hyperref}
\usepackage{mathrsfs}

\usepackage{enumerate}
\usepackage{ulem}
\usepackage{afterpage}

\hypersetup{%
 setpagesize=false,
 bookmarksnumbered=true,%
 bookmarksopen=true,%
 colorlinks=true,%
 linkcolor=blue,%
 citecolor=red}

\begin{document}




\title{
Multi-rotating black holes with non-aligned angular momenta \\in  5D Kaluza-Klein theory
}

\author{Shinya Tomizawa$^1$}
\email{tomizawa@toyota-ti.ac.jp}

\author{Hideo Furugori$^1$}
\email{hideo@toyota-ti.ac.jp}


\author{Jun-ichi Sakamoto$^2$}
\email{jsakamoto@het.phys.sci.osaka-u.ac.jp}

\author{ Ryotaku Suzuki$^3$}
\email{suzuki.ryotaku@nihon-u.ac.jp}

\affiliation{\vspace{5mm}
$^{1}$ Mathematical Physics Laboratory, Toyota Technological Institute, Hisakata 2-12-1, Tempaku-ku, Nagoya, Japan 468-8511
\vspace{5mm}\\
$^{2}$Department of Physics, The University of Osaka
Machikaneyama-Cho 1-1, Toyonaka, Japan 560-0043,
\vspace{5mm}\\
$^3$Laboratory of Physics, College of Science and Technology, Nihon University, Narashinodai 7-24-1, Funabashi, Chiba 274-8501, Japan}

\date{\today}

\preprint{TTI-MATHPHYS-43}
\preprint{OU-HET-1318}




\begin{abstract} 
We present an exact solution describing multi-rotating black holes in 4D Einstein-Maxwell-dilaton theory, which can be obtained from 5D Kaluza--Klein theory via dimensional reduction.
The solution represents a multi-centered configuration of rotating black holes carrying both electric and magnetic charges, with each black hole possessing a non-aligned angular momentum.
This work generalizes our previous solution for black holes with aligned angular momenta to the more general case of non-aligned angular momenta.
It includes, as special cases, the Majumdar--Papapetrou solution, the recent multi-centered rotating black hole solutions of Teo and Wan, and our previous solution with unequal electric and magnetic charges.
The resulting spacetimes are free of curvature singularities and closed timelike curves, both on and outside the horizons, provided that the magnitude of the spin angular momentum of each black hole remains below a certain upper bound.
\end{abstract}

\date{\today}
\maketitle



\section{Introduction}

Gravitational solutions of multi-black-hole systems are of fundamental importance in both astrophysics and theoretical physics. 
In particular, they provide valuable insights into the nonlinear interaction of compact objects, which play a central role in gravitational wave physics. 
Despite their importance, constructing such solutions remains highly nontrivial, mainly due to the absence of sufficient symmetries and the intrinsically nonlinear nature of the Einstein equations.
In vacuum general relativity, it has long been known that static or stationary multi-black-hole configurations generically suffer from singular structures. 
A classical example is the Israel--Khan solution~\cite{Israel1964}, which describes multiple Schwarzschild black holes aligned along a common axis. 
In this configuration, the gravitational attraction between the black holes must be balanced by conical singularities along the axis. 
This feature persists even in rotating generalizations, such as the double Kerr solution constructed by Kramer and Neugebauer~\cite{Kramer1980}, where the interplay between gravitational attraction and spin--spin interaction still fails to eliminate conical singularities. 
Tan and Teo~\cite{Tan:2003jz} constructed static multi-black hole solutions in 5D vacuum Einstein gravity as higher-dimensional generalizations of the Israel-Khan solution. These configurations describe multiple Schwarzschild black holes aligned along a common axis within the Weyl formalism. However, the solutions necessarily contain conical singularities along the axis, indicating that gravitational attraction cannot be balanced in vacuum. 
This shows that regular static multi-black-hole equilibrium configurations do not exist in pure Einstein gravity.
Herdeiro et al.~\cite{Herdeiro:2008en} obtained a solution describing two rotating Myers-Perry black holes by using the inverse scattering method. This provides an explicit example of interacting rotating black holes in 5D vacuum gravity.
Although spin-spin interactions introduce repulsive effects, they are insufficient to achieve equilibrium, and conical singularities persist. This confirms that rotation alone does not remove singular structures in multi-black-hole systems. 
These results strongly suggest that regular multi-black-hole solutions do not exist in asymptotically flat vacuum spacetimes. 
Recently, we have presented new families of multi-centered extremal Myers-Perry black hole solutions with asymptotically locally Euclidean (ALE) asymptotics in 5D vacuum Einstein gravity~\cite{Tomizawa:2026soz}.  
Remarkably, these solutions are free of curvature singularities outside the horizons, admit no closed timelike curves, and contain no conical singularities between the black holes.

\medskip
In Newtonian gravity, a system of charged point particles can remain in static equilibrium when the charges $Q_i$ (of the same sign) are equal to the masses $M_i$, so that the electrostatic repulsion precisely balances the gravitational attraction. This balance condition, often referred to as the BPS condition $Q_i = M_i$, admits a natural generalization to general relativity.
In 4D Einstein--Maxwell theory, the Majumdar--Papapetrou solution~\cite{Majumdar:1947eu,Papapetrou} provides the first explicit realization of such equilibrium configurations in terms of multiple extremal black holes. Since each constituent satisfies the BPS condition, the net force between black holes vanishes, allowing for static multi-black-hole configurations without conical singularities. However, this mechanism relies crucially on extremality and does not extend straightforwardly to rotating systems. Indeed, stationary generalizations constructed by Israel, Wilson, and Perj\'es~\cite{Perjes:1971gv,Israel:1972vx} were later shown to describe naked singularities rather than regular black holes. More recently, related multi-black-hole configurations have also been constructed in non-asymptotically flat backgrounds, such as the Bertotti--Robinson spacetime~\cite{DiPinto:2026rvp,Furugori:2026pdt}. 
These constructions have been extended to higher dimensions. 
In particular, Myers~\cite{Myers:1986rx} generalized the Majumdar--Papapetrou solutions to arbitrary spacetime dimensions $D \geq 5$, providing a class of extremal multi-black-hole solutions in higher-dimensional Einstein--Maxwell theory. 
In 5D minimal supergravity, which corresponds to 5D Einstein-Maxwell-Chern-Simons theory with a specific coupling constant, the BMPV black hole~\cite{Breckenridge:1996is} represents a supersymmetric rotating solution with a regular $S^3$ horizon. 
Multi-centered generalizations of this BMPV solution were subsequently developed in~\cite{Gauntlett:1998fz}, where a wide class of BPS configurations can be described in terms of harmonic functions on a 3D base space.

\medskip
More recently, significant progress has been made in higher-dimensional theories. 
In particular, Teo and Wan~\cite{Teo:2023wfd} constructed a class of  non-BPS solutions describing multi-centered rotating black holes in 5D Kaluza--Klein theory. 
Under dimensional reduction along the fifth spatial dimension, these solutions describe balanced configurations of rotating dyonic black holes with equal electric and magnetic charges  in 4D Einstein--Maxwell--dilaton theory. 
Remarkably, these solutions are free from curvature singularities, conical defects, Dirac--Misner strings, and closed timelike curves. 
Moreover, the previous work~\cite{Tomizawa:2025tvb}  extends the recent multi-centered rotating black hole solutions of Teo and Wan to configurations with unequal electric and magnetic charges. 
The resulting spacetimes are free of curvature singularities, conical defects, Dirac-Misner strings, and closed timelike curves, both on and outside the horizons, provided that the black holes have either aligned or anti-aligned spin orientations.
However, in these solutions, the restriction to aligned or antialigned angular momenta is a strong limitation. 
From both physical and geometrical viewpoints, it is natural to ask whether regular multi-black-hole configurations can exist with more general spin orientations. 
In particular, allowing non-aligned angular momenta breaks axial symmetry and significantly complicates the structure of the spacetime, making the construction of exact solutions highly nontrivial.
In this paper, we overcome this difficulty and construct a new class of exact solutions describing multi-rotating black holes with non-aligned angular momenta in 5D Kaluza--Klein theory. Our solution generalizes the Teo--Wan configuration to the case of fully non-aligned spins and allows for arbitrary orientations of the individual angular momenta. The resulting spacetime is asymptotically flat and remains completely regular on and outside the horizons, 
without curvature singularities, or closed timelike curves, 
provided that appropriate bounds on the angular momenta are satisfied.
The key feature of our construction is that it retains a description in terms of harmonic functions on three-dimensional Euclidean space, despite the absence of axial symmetry. This provides a novel mechanism for constructing regular multi-black-hole configurations beyond the aligned case and offers a new perspective on the interplay between rotation, charge, and regularity in higher-dimensional gravity.

\medskip

The organization of this paper is as follows. In Sec.~\ref{sec:our}, we review the multi-centered rotating black hole solutions presented in our previous work~\cite{Tomizawa:2025tvb}. This solution describes black holes whose angular momenta are aligned or anti-aligned with the $z$-axis on a 3D flat space.
In Sec.~\ref{sec:formalism}, we outline the nonlinear sigma model formulation based on the work of Maison and Cl\'ement. 
The previous solution can be expressed in terms of a matrix exponential involving two harmonic functions defined on a 3D flat space: one corresponds to a monopole potential associated with masses or charges, while the other represents a dipole potential associated with aligned angular momenta.
In Sec.~\ref{sec:solution}, we construct a new solution by extending the harmonic functions to multiple centers with non-aligned angular momenta.
In Sec.~\ref{sec:analysis}, we analyze the properties of the solution, including its regularity, asymptotic behavior, and the absence of closed timelike curves. As a simple illustration, we also discuss the special case of two black holes.
Finally, Sec.~\ref{sec:summary} is devoted to a summary and discussion.

\section{Our previous solutions}\label{sec:our}

In this section, we review our previous multi-centered rotating black hole solutions in 5D Kaluza--Klein theory~\cite{Tomizawa:2025tvb}, which generalize the equal-charge solutions of Teo and Wan~\cite{Teo:2023wfd} to the case of unequal charges.
We begin with 5D Kaluza--Klein theory, in which the metric takes the form
\begin{eqnarray}
ds^2=e^{-\frac{2\phi}{\sqrt{3}}}(dx^5+{\bm A})^2+e^{\frac{\phi}{\sqrt{3}}}g_{\mu\nu} dx^\mu dx^\nu,
\end{eqnarray}
where the scalar field $\phi$, the components $A_\mu$ of the one-form ${\bm A}=A_\mu dx^\mu$, and the four-dimensional metric $g_{\mu\nu}$ ($\mu,\nu=0,\ldots,3$) are independent of the compact coordinate $x^5$, which is assumed to have period $2\pi R_{KK}$.
Upon dimensional reduction, the 5D  Einstein theory gives rise to the 4D  Einstein--Maxwell--dilaton system, described by the action: 
\begin{eqnarray}
S=\int d^4x\sqrt{-g} \left( R-\frac{1}{2}\partial_\mu \phi\partial^\mu \phi -\frac{1}{4}e^{-\sqrt{3} \phi} F^2 \right),
\end{eqnarray}
where $R$ denotes the Ricci scalar associated with $g_{\mu\nu}$, $g={\rm det}(g_{\mu\nu})$, and $F_{\mu\nu}:=\partial_\mu A_\nu-\partial_\nu A_\mu$ is the field strength.
The corresponding equations of motion consist of the Einstein equation for $g_{\mu\nu}$, together with the equations for the gauge field $A_\mu$ and the dilaton $\phi$, which take the form
\begin{eqnarray}
R_{\mu\nu}=\frac{1}{2}\partial_\mu \phi \partial_\nu \phi +\frac{1}{2}e^{-\sqrt{3} \phi} \left(F_{\mu\rho}F_\nu{}^\rho-\frac{1}{4} g_{\mu\nu} F^2 \right),
\end{eqnarray}
\begin{eqnarray}
\nabla_\mu\left(e^{-\sqrt{3} \phi}F^{\mu\nu}\right)=0,
\end{eqnarray}
\begin{eqnarray}
\nabla_\mu\nabla^\mu\phi=-\frac{\sqrt{3}}{4} e^{-\sqrt{3} \phi} F^2.
\end{eqnarray}

The 5D metric for the multi-centered black hole solution takes the form
\begin{eqnarray}
ds^2=\frac{H_-}{H_+}\left[dx^5-\frac{dt}{\sin\alpha}+\frac{1+f \sin^22\alpha}{\sin\alpha\  H_-}(dt+{\bm \omega^0})+\tilde {\bm \omega}^5\right]^2-\frac{1}{H_-}(dt+{\bm \omega^0})^2+H_+d{\bm x}\cdot d{\bm x},
\end{eqnarray}
and the 4D metric, gauge field and scalar field after dimensional reduction are expressed as
\begin{eqnarray}
ds^2_{(4)}&=&-\frac{1}{\sqrt{H_+H_-}}(dt+{\bm \omega^0})^2+\sqrt{H_+H_-}d{\bm x}\cdot d{\bm x}, \label{eq:sol:4Dmetric}\\ 
{\bm A}&=&\left(-\frac{1}{\sin\alpha}+\frac{1+f \sin^22\alpha}{\sin\alpha\  H_-}\right)dt+
\frac{1+f \sin^22\alpha}{\sin\alpha\  H_-}{\bm \omega^0}+\tilde {\bm \omega}^5, \label{eq:sol:A}\\
e^{\frac{2\phi}{\sqrt{3}}}&=&\frac{H_+}{H_-},\label{eq:sol:phi}
\end{eqnarray}
where $d{\bm x}\cdot d{\bm x}$ is the metric of the 3D Euclid space ${\mathbb E}^3$ with ${\bm x}=(x,y,z)$. 
The functions $H_\pm$, one-forms ${\bm \omega}^0$, $\tilde{\bm \omega}^5$ on ${\mathbb E}_3$ are given by
\begin{eqnarray}
H_\pm&=&1+2f+\sin^22\alpha\ f^2+2\sin^22\alpha \cos2\alpha\ g\pm (2\sin^22\alpha \ g+2\cos 2\alpha\ f + \cos 2\alpha \sin^2 2\alpha\ f^2),\label{eq:Hpma}\\
{\bm \omega}^0&=&-\sum_{i=1}^N\frac{2 J_i [(y-y_i)dx-(x-x_i)dy]}{|{\bm x}-{\bm x}_i|^3}, \label{eq:omega0_a}\\
\tilde{\bm \omega}^5&=& \sum_{i=1}^N\frac{2P_i(z-z_i)}{|{\bm x}-{\bm x_i}|}\frac{(y-y_i)dx-(x-x_i)dy}{(x-x_i)^2+(y-y_i)^2} , \label{eq:omega5_a}
\end{eqnarray}
with the two harmonic functions, $f$ and $g$, having point sources at the positions ${\bm x}={\bm x_i}:=(x_i,y_i,z_i)$ ($i=1,\ldots,N$) on ${\mathbb E}^3$,
\begin{eqnarray}
f=\sum_{i=1}^N\frac{M_i}{|{\bm x}-{\bm x_i}|},\quad 
g=\sum_{i=1}^N\frac{J_iM_i^2(z-z_i)}{2P_iQ_i|{\bm x}-{\bm x_i}|^3}, \label{eq:gma}
\end{eqnarray}
This solution describes an asymptotically flat, stationary multi-centered rotating dyonic black holes, with each having an extremal horizon.
The $i$-th black hole at the position ${\bm x}={\bm x}_i$ on ${\mathbb E}^3$ carries the mass $M_i$, spin angular momentum $J_i$, electric and magnetic charges  $Q_i$ and $P_i$, respectively, given by 
\begin{eqnarray}
\frac{P_i}{2M_i}=\cos^3\alpha,\quad \frac{Q_i}{2M_i}=\sin^3\alpha\ (i=1,2,\ldots N). \label{eq:PiQi}
\end{eqnarray}
From Eqs.~(\ref{eq:omega0_a}) and (\ref{eq:gma}), we see that each black hole has either an aligned or anti-aligned spin orientation along the $z$-axis.
Furthermore, the regularity of the metric on the horizon requires the  condition
\begin{eqnarray}
|J_i|<P_iQ_i, \label{eq:regularity}
\end{eqnarray}
since the horizon area, 
$8\pi \sqrt{P_i^2Q_i^2-J_i^2}$, 
vanishes if this is saturated. 
Moreover, as shown in Ref.~\cite{Teo:2023wfd}, under the condition~(\ref{eq:regularity}), the spacetime is free from closed timelike curves (CTCs) on and outside the horizon.

\medskip
In the case of equal charges, $P_i = Q_i = M_i/\sqrt{2}$ (i.e., $\alpha = \pi/4$), and in the limit $J_i \to 0$ for all $i$, the scalar field $\phi$ vanishes. Consequently, the Majumdar--Papapetrou solution describing static multiple dyonic black holes is recovered.

\section{5D Kaluza-Klein theory and non-linear sigma model}\label{sec:formalism}
Since we wish to consider stationary solutions, in addition to the Killing vector $\partial/\partial x^5$, we assume the existence of a timelike Killing vector field $\partial/\partial t$. With these two commuting Killing vectors, the 5D Einstein equations reduce to a 3D system of gravity coupled to scalar fields~\cite{Maison:1979kx}. We then review how this system of scalar fields can be described by a nonlinear sigma model. In general, solving the resulting equations is challenging because the scalar fields are coupled to 3D gravity. However, under the assumption that the 3D geometry is flat, Cl\'ement~\cite{Clement:1985gm,Clement:1986bt} showed that a special class of solutions can be constructed using two harmonic functions. We also present the necessary equations for our analysis, following Cl\'ement's formulation.

\subsection{5D Einstein gravity with two commuting Killing vectors}

Let $\xi_a \ (a=0,5)$ denote two commuting Killing vector fields, satisfying $[\xi_a,\xi_b]=0$ and ${\cal L}_{\xi_a} g=0$. Introducing coordinates $x^a$ adapted to these Killing vectors, such that $\xi_a = \partial/\partial x^a$, the metric can be written in the form
\begin{eqnarray}
ds^2 = \lambda_{ab}(dx^a+\omega^a{}_{i}dx^i)(dx^b+\omega^b{}_{j}dx^j)
+|\tau|^{-1}h_{ij}dx^idx^j , \label{eq:5D metric}
\end{eqnarray}
where $\tau := -\det(\lambda_{ab})$, and the quantities $\omega^a{}_{i}$ and $h_{ij}$ ($i=1,2,3$) are independent of the coordinates $x^a$.
We further define the twist one-forms by
\begin{equation}
v_a = *(\xi_0 \wedge \xi_5 \wedge d\xi_a) .
\end{equation}
Their exterior derivatives are given by
\begin{eqnarray}
dv_a = 2 *(\xi_0 \wedge \xi_5 \wedge R(\xi_a)) ,
\end{eqnarray}
where $R(\xi_a)$ denote the Ricci one-forms.

\medskip
Since $R(\xi_a)=0$ by virtue of the vacuum Einstein equations, there exist locally defined twist potentials $V_a$ satisfying
\begin{eqnarray}
 dV_a=v_a,
\label{eq:twistpotential} 
\end{eqnarray} 
which can be written as
\begin{eqnarray}
  \partial_kV_{a}
 =\tau \sqrt{|h|} \lambda_{ab} \varepsilon_{kij}  h^{im}h^{jn}\partial_m \omega^b{}_n.\label{eq:domega}
\end{eqnarray}
The vacuum Einstein equations can then be reformulated as the field equations for the five scalar fields, ${\lambda_{ab},V_a}$,
\begin{eqnarray}
\Delta_h \lambda_{ab}&=&\lambda^{cd} h^{ij}\frac{\partial \lambda_{ac}}{\partial x^i} \frac{\partial \lambda_{bd}}{\partial x^j}+\tau^{-1} h^{ij}\frac{\partial V_a}{\partial x^i} \frac{\partial V_b}{\partial x^j},\label{eq:eom1}\\
\Delta_h V_{a}&=& \tau^{-1} h^{ij}\frac{\partial \tau}{\partial x^i} \frac{\partial V_a}{\partial x^j}+\lambda^{bc} h^{ij}\frac{\partial \lambda_{ab}}{\partial x^i} \frac{\partial V_c}{\partial x^j},\label{eq:eom2}
\end{eqnarray}
and the Einstein equations for the 3D metric $h_{ij}$, which is coupled with the five scalar fields,
\begin{eqnarray}
R^h_{ij} &=&  \frac{1}{4} \lambda^{ab}\lambda^{cd}
              \frac{\partial \lambda_{ac}}{\partial x^i }  \frac{\partial \lambda_{bd}}{\partial x^j } 
   + \frac{1}{4}\tau^{-2}\frac{\partial \tau}{\partial x^i} \frac{\partial \tau}{\partial x^j }   
    -\frac{1}{2}\tau^{-1}\lambda^{ab} \frac{\partial V_a}{\partial x^i }\frac{\partial V_b}{\partial x^j },       
\label{eq:Rij}
\end{eqnarray} 
where $\Delta_h$ denotes the Laplacian with respect to $h_{ij}$, and $R^h_{ij}$ is the corresponding Ricci tensor.

\medskip
\subsection{$SL(3,{\mathbb R})$ nonlinear sigma model}
As a consequence of the presence of two isometries, the system is described by five scalar fields $\lambda_{ab}$ and $V_a$ $(a=0,5)$, which we collectively denote by $\Phi^A=(\lambda_{ab},V_a)$. It can be shown that the equations of motion, Eqs.~(\ref{eq:eom1}), (\ref{eq:eom2}) and (\ref{eq:Rij}), follow from the following action for the sigma model $\Phi^A$ coupled to 3D gravity with respect to the metric $h_{ij}$:
\begin{eqnarray}
S=\int\left(R^h
 -G_{AB}\frac{\partial \Phi^A}{\partial x^i}
 \frac{\partial \Phi^B}{\partial x^j}h^{ij}\right)\sqrt{|h|}d^3x \,,\label{eq:action}
\end{eqnarray}
where the target space metric $G_{AB}$ is given by 
\begin{eqnarray}
G_{AB}d\Phi^Ad\Phi^B 
&=& \frac{1}{4}{\rm Tr}(\lambda^{-1}d\lambda\lambda^{-1}d\lambda )
   + \frac{1}{4}\tau^{-2}d\tau^2 
    -\frac{1}{2}\tau^{-1}v^T\lambda^{-1}v,
\end{eqnarray}
with $\lambda=(\lambda_{ab})$, $v=(v_0,v_5)^T$ and $v=dV$.
Varying the action with respect to $h_{ij}$ yields ~(\ref{eq:Rij}),
\begin{eqnarray}
R^h_{ij} &=& G_{AB}\frac{\partial \Phi^A}{\partial x^i} 
                 \frac{\partial \Phi^B}{\partial x^j},  
      \end{eqnarray} 
while variation with respect to $\Phi^A$ leads to Eqs.~(\ref{eq:eom1}) and (\ref{eq:eom2}),
\begin{eqnarray}
\Delta_h\Phi^A+h^{ij}\Gamma^A_{BC}\frac{\partial \Phi^B}{\partial x^i} 
                                  \frac{\partial \Phi^C}{\partial x^j} 
  =0, \label{eq:harmonic_map}
\end{eqnarray}
where $\Gamma^A_{BC}$ denotes the Christoffel symbols associated with the metric $G_{AB}$.

\subsection{Coset matrix}

It was shown by Maison~\cite{Maison:1979kx} that the action~(\ref{eq:action})  describing the five scalar fields $\Phi^A$ possesses a global $SL(3,\mathbb{R})$ symmetry. This structure can be made manifest by introducing the $SL(3,\mathbb{R})$-valued matrix $\chi$, defined by the $3\times3$ matrix:
\begin{eqnarray}
\chi= \left(
  \begin{array}{ccc}
  \displaystyle \lambda_{ab}-\frac{V_ aV_b^T}{\tau} &  \displaystyle \frac{V_a}{\tau}\\
   \displaystyle  \frac{V_b^T}{\tau}                     & \displaystyle  -\frac{1}{\tau}
    \end{array}
 \right) \,, \label{eq:chidef}
\end{eqnarray}
which satisfies $\chi^T=\chi$ and $\det(\chi)=1$.
With this definition, the action~(\ref{eq:action}) takes the form:
\begin{eqnarray}
S=\int\left(R^h
 - \frac{1}{4}h^{ij} {\rm tr}(\chi^{-1}\partial_i\chi \chi^{-1}\partial_j\chi)
\right)\sqrt{|h|}d^3x \,,
\end{eqnarray}
which  is invariant under the transformation
\begin{eqnarray}
\chi\to \chi'=g \chi g^T,\quad h\to h \label{eq:sl3r}
\end{eqnarray}
with $g\in SL(3,{\mathbb R})$.
In this formulation, the equations of motion~(\ref{eq:Rij}), (\ref{eq:eom1}) and  (\ref{eq:eom2})  reduce to
\begin{eqnarray}\label{eq:eom}
&&d\star_h (\chi^{-1} d\chi)=0, \label{eq:eomb}\\
&&R^h_{ij}=\frac{1}{4}{\rm tr}(\chi^{-1}\partial_i\chi \chi^{-1}\partial_j\chi).\label{eq:eom2b}
\end{eqnarray}
which provide a compact representation.
In summary, the presence of two commuting Killing vectors allows the 5D vacuum Einstein system to be recast as a 3D  nonlinear sigma model with $SL(3,\mathbb{R})$ symmetry. The precise coset structure depends on the causal character of the Killing vectors: if both are spacelike, the target space is $SL(3,\mathbb{R})/SO(3)$, whereas if one is timelike, it becomes $SL(3,\mathbb{R})/SO(2,1)$.

\subsection{A class of 4D asymptotically flat solutions}
Because Eqs.~(\ref{eq:eomb}) and (\ref{eq:eom2b}) are coupled to each other, solving them in general is a nontrivial task. Nevertheless, a significant simplification occurs when the three-dimensional metric $h_{ij}$ is taken to be the flat Euclidean metric on ${\mathbb E}^3$,
\begin{eqnarray}
h_{ij}dx^idx^j=d{\bm x}\cdot d{\bm x}, \label{eq:h}
\end{eqnarray}
where ${\bm x}=(x,y,z)$ denotes the position vector on ${\mathbb E}^3$. In this case, the equations reduce to
\begin{eqnarray}
&&\partial_i (\chi^{-1} \partial^i\chi)=0,\label{eq:eomc}\\
&&{\rm tr}(\chi^{-1}\partial_i\chi \chi^{-1}\partial_j\chi)=0.\label{eq:eom2c}
\end{eqnarray}
A general class of solutions depending on two scalar potentials was obtained by
Cl\'ement~\cite{Clement:1986bt,Clement:1985gm}, and can be written as
\begin{eqnarray}
\chi=\eta e^{fA}e^{gA^2}, \label{eq:chi}
\end{eqnarray}
where $f$ and $g$ are harmonic functions defined on ${\mathbb E}^3$, and $\eta$ and $A$ are constant $3\times 3$ matrices.
To ensure asymptotic flatness of the four-dimensional metric $ds^2_{(4)}=g_{\mu\nu}dx^\mu dx^\nu$, the matrix $\eta$ is chosen as
\begin{eqnarray}
\eta=
\begin{pmatrix}
-1  &  0& 0 \\
  0&1 &  0   \\
0  & 0 &-1  
\end{pmatrix},
\end{eqnarray}
provided that the harmonic functions vanish at spatial infinity.
Furthermore, if the matrix $A$ satisfies
\begin{eqnarray}
A^T=\eta A \eta, \quad {\rm tr}(A)=0, \quad 
{\rm tr}(A^2)=0,
\end{eqnarray}
then the matrix $\chi$ automatically satisfies the required conditions: it is symmetric ($\chi^T=\chi$), unimodular (${\rm det}\ \chi=1$), and obeys the constraint~(\ref{eq:eom2c}).

\section{Multi-centered rotating black holes with non-aligned angular momenta}\label{sec:solution}

An exact solution describing the most general dyonic rotating black holes in 5D Kaluza--Klein theory was obtained by Rasheed and Larsen~\cite{Rasheed,Larsen:1999pp}. This solution exhibits two qualitatively different extremal limits, corresponding to the slow-rotation and fast-rotation regimes. Among these, only the slowly rotating limit can be accommodated within Cl\'ement's framework, while the fast-rotating case cannot be captured in this way. In the former case, the solution admits a description in terms of two harmonic functions, as in Eq.~(\ref{eq:chi}), each arising from a single point  source.
In our previous work~\cite{Tomizawa:2025tvb}, by generalizing these harmonic functions to multiple point sources, we constructed an exact solution describing multi-centered rotating dyonic black holes with unequal electric and magnetic charges and aligned (or anti-aligned) angular momenta in 4D Einstein--Maxwell--dilaton theory. 
This extends the construction of Ref.~\cite{Teo:2023wfd}, where the charges were taken to be equal. 
In this section, we generalize our previous solution for black holes with aligned (or anti-aligned) angular momenta to the case of non-aligned angular momenta.

\subsection{Construction of our previous solutions}

\medskip

Our previous solution belongs to the class of solutions~(\ref{eq:chi}), characterized by two harmonic functions introduced by Cl\'ement. The corresponding functions $f$ and $g$ are given by (\ref{eq:gma}), and the matrix $A$ is given by
\begin{eqnarray}
A=4
\begin{pmatrix}
   -s^2 &  s^3 & 0 \\
  - s^3& s^2-c^2 &   c^3  \\
  0 &  -c^3&    c^2
\end{pmatrix}
,\label{eq:A}
\end{eqnarray}
where $(c,s):=(\cos\alpha,\sin\alpha)$, and the parameter $\alpha$ characterizes the ratio of electric to magnetic charges.

\begin{eqnarray}
\setlength{\arraycolsep}{1pt}  
\chi
=
{\tiny
\begin{pmatrix}
 -8 s^4 c^2 f^2+4 s^2 f-16 c^6 g+32 c^4 g-16 c^2 g-1 & 4 s^3 \left(2 c^2 f^2+4 c^2 g-f\right) & -8 c^3s^3\left(f^2+2 g\right) \\
 4s^3\left(2 c^2 f^2+4 c^2 g-f\right) & -8s^2 c^2 f^2+\left(4-8 c^2\right) f+16 c^4 g-16 c^2 g+1 & 4 c^3 \left(2 s^2f^2+4 s^2 g+f\right) \\
 -8 c^3s^3\left(f^2+2 g\right) & 4 c^3 \left(2 s^2f^2+4s^2 g+f\right) & 8 c^6 \left(f^2+2 g\right)-8 c^4 \left(f^2+2 g\right)-4 c^2 f-1 \\
\end{pmatrix}
}.
\end{eqnarray}
Hence, from Eq.~(\ref{eq:chidef}), we can read off the conformal factor $\tau$ and the scalar fields $(\lambda_{ab},V_a)$ as
\begin{eqnarray}
\tau&=&\frac{1}{8 s^2c^4 \left(f^2+2 g\right)+4 c^2 f+1},\\
\lambda_{00}&=&-\tau[-8 s^2 c^2 f^2+\left(8 c^2-4\right) f+16c^2s^2g+1],\\
\lambda_{05}&=&-4\tau s^3 \left(2 c^2 f^2+f-4 c^2 g\right),\\
\lambda_{55}&=&-\tau \left[-8 s^4 c^2 f^2-4s^2 f+16c^2s^4g-1\right] ,\\
V_0&=& -8\tau c^3 s^3 \left(f^2+2 g\right), \label{eq:V0}\\
V_5&=& 4c^3\tau( 2 s^2 f^2 + f +4 s^2g). \label{eq:V5}
\end{eqnarray}
From Eqs.~(\ref{eq:V0}) and (\ref{eq:V5}), the 1-forms ${\bm \omega}^0$ and ${\bm \omega}^5$ can be expressed as
\begin{eqnarray}
{\bm \nabla} \times {\bm \omega}^0&=&-2\sin^32\alpha {\bm \nabla} g \label{eq:omega0a},\\
{\bm \nabla} \times {\bm \omega}^5&=&-4\cos^3\alpha ({\bm \nabla} f+4\sin^2\alpha {\bm \nabla}g ).\label{eq:omega5}
\end{eqnarray}
If we define $\tilde {\bm \omega}^5:={\bm \omega^5}-(\sin\alpha)^{-1} \ {\bm \omega^0}$, then
\begin{eqnarray}
{\bm \nabla} \times \tilde {\bm \omega}^5=-4\cos^3\alpha {\bm \nabla}  f.  \label{eq:omega5t}
\end{eqnarray}
From Eq.~(\ref{eq:gma}),  these can be solved as  (\ref{eq:omega0_a}) and (\ref{eq:omega5_a}).

\subsection{Multi-centered rotating black holes with non-aligned angular momenta}

We now consider the harmonic functions $f$ and $g$ in Eq.~(\ref{eq:gma}) and generalize them to the following multi-centered form:
\begin{eqnarray}
f&=&\sum_{i=1}^N\frac{M_i}{|{\bm x}-{\bm x_i}|},  \label{eq:fm}\\
g&=&\sum_{i=1}^N\frac{M_i^2{\bm J_i}\cdot ({\bm x}-{\bm x_i}) }{2P_iQ_i |{\bm x}-{\bm x_i}|^3}. \label{eq:gm}
\end{eqnarray}
Here, the constant vector ${\bm J_i}=(J_{x,i},J_{y,i},J_{z,i})$ $(i=1,\ldots,N)$ denotes the angular momentum of each black hole.  
In our previous work~\cite{Tomizawa:2026soz}, all angular momenta ${\bm J_i}$ were aligned or  anti-aligned  along the $z$-axis, i.e., $J_{x,i}=J_{y,i}=0$. 
In contrast, the present solution allows for non-aligned angular momenta, so that their directions are arbitrary.

\medskip
Using  Eqs.~(\ref{eq:omega0a}) and (\ref{eq:omega5t}), we can show that the corresponding one-forms ${\bm \omega}^0$ and $\tilde {\bm \omega}^5$ take the following explicit forms:
\begin{eqnarray}
{\bm \omega}^0&=&\sum_{i=1}^N\frac{2{\bm J_i}\cdot[({\bm x}-{\bm x_i})\times d{\bm x}]}{|{\bm x}-{\bm x}_i|^3}, \label{eq:omega0}\\
 \tilde{\bm \omega}^5&=& \sum_{i=1}^N\frac{2P_i(z-z_i)}{|{\bm x}-{\bm x_i}|}\frac{(y-y_i)dx-(x-x_i)dy}{(x-x_i)^2+(y-y_i)^2}.\label{eq:tomega5}
\end{eqnarray}
Consequently,  the 5D metric for the multi-centered black hole solution takes the form
\begin{eqnarray}
ds^2=\frac{H_-}{H_+}\left[dx^5-\frac{dt}{\sin\alpha}+\frac{1+f \sin^22\alpha}{\sin\alpha\  H_-}(dt+{\bm \omega^0})+\tilde {\bm \omega}^5\right]^2-\frac{1}{H_-}(dt+{\bm \omega^0})^2+H_+d{\bm x}\cdot d{\bm x},
\end{eqnarray}
with Eqs.~(\ref{eq:Hpma}), (\ref{eq:fm}), (\ref{eq:gm}), (\ref{eq:omega0}), and (\ref{eq:tomega5}).
\medskip
On the other hand, the 4D metric, gauge field and scalar field after dimensional reduction are expressed as
\begin{eqnarray}
ds^2_{(4)}&=&-\frac{1}{\sqrt{H_+H_-}}(dt+{\bm \omega^0})^2+\sqrt{H_+H_-}d{\bm x}\cdot d{\bm x}, \label{eq:sol:4Dmetric}\\ 
{\bm A}&=&\left(-\frac{1}{\sin\alpha}+\frac{1+f \sin^22\alpha}{\sin\alpha\  H_-}\right)dt+
\frac{1+f \sin^22\alpha}{\sin\alpha\  H_-}{\bm \omega^0}+\tilde {\bm \omega}^5, \label{eq:sol:A}\\
e^{\frac{2\phi}{\sqrt{3}}}&=&\frac{H_+}{H_-}.\label{eq:sol:phi}
\end{eqnarray}

\section{Properties of the multi-rotating black hole solution}\label{sec:analysis}

In this section, we see that this solution is regular and describes asymptotically flat, multicentered rotating dyonic black holes, each possessing an extremal horizon.
We demonstrate that curvature singularities are confined inside the horizons and do not occur on or outside them.
Furthermore, we prove the absence of CTCs in the exterior region as well as on the horizons.

\subsection{Near-horizon geometry}

The 4D metric appears to diverge at the points ${\bm x}={\bm x}_i$. However, we show that these points correspond to smooth Killing horizons, provided the slow-rotation condition $|{\bm J}_i|<|P_iQ_i|$ holds for all $i$.
We now focus on the $i$-th center and show that it corresponds to a Killing horizon with spherical topology. Introducing $r:=|{\bm x}-{\bm x}_i|$ and standard spherical coordinates $(x,y,z)=(r\sin\theta \cos\phi, r\sin\theta\sin\phi, r\cos\theta)$, we may, without loss of generality, align the $z$-axis with the angular momentum vector, so that ${\bm J}_i=(0,0,J_i)$.
Under this choice, we consider the limit $r\to 0$. In this limit, the 4D metric behaves as

\begin{eqnarray}
ds^2_{(4)}&\simeq&-\frac{ r^2}{2\sqrt{P_i^2Q_i^2 -J_i^2 \cos^2\theta }}\left[dt+\frac{2J_i}{r}\sin^2\theta d\phi \right]^2+2\sqrt{P_i^2Q^2_i-J_i^2\cos^2\theta } \left[ \frac{dr^2}{r^2}+ d\Omega^2 \right],
\end{eqnarray}
The metric component $g_{rr}$ evidently diverges at $r=0$, however, this divergence is only apparent, as we now demonstrate. 
Introducing new coordinates $(v,\phi')$ defined by
\begin{eqnarray}
dt=dv+\left(\frac{a_0}{r^2}+\frac{a_1}{r} \right)dr,\quad d\phi=d\phi'+\frac{b_0}{r}dr,
\end{eqnarray}
we can choose the coefficients so as to eliminate the divergences in the metric components,  $g_{r\phi'}=O(r^{-1})$ and $g_{rr}=O(r^{-2})$, 
as 
\begin{eqnarray}
a_0&=&\pm2\sqrt{P_i^2Q_i^2-J_i^2},\\
a_1&=&\frac{2M_i^3\sin^42\alpha[1+(\sum_{j\not=i}M_j|{\bm x}_j|^{-1})\sin^22\alpha ]}{a_0},\\
b_0&=&\frac{2J_i}{a_0}.
\end{eqnarray}

\medskip
This means that the null hypersurface at $r=0$ constitutes a Killing horizon generated by the vector field $\partial/\partial v$. To analyze its geometry, we introduce the near-horizon scaling $v \to v/\varepsilon$, $r \to \varepsilon r$, and take the limit $\varepsilon \to 0$. In this limit, the metric takes the form
\begin{eqnarray}
ds^2_{(4)}&\simeq&\frac{2f(0)^2\sin^2\theta}{f(\theta)}\left[d\phi'-\frac{rJ_i}{2f(0)^2}dv \right]^2
-\frac{f(\theta)r^2}{2f(0)^2}dv^2
\mp\frac{2f(\theta)}{f(0)}dvdr+2f(\theta)d\theta^2,
\end{eqnarray}
where $f(\theta):=\sqrt{P_i^2Q_i^2-J_i^2\cos^2\theta}$. This metric matches the near-horizon geometry of a singly rotating extremal Rasheed--Larsen black hole in the slow-rotation extremal limit.

\subsection{Asymptotic structure}

At infinity $r \to \infty$, the 4D fields behave as follows:
\begin{eqnarray}
ds^2_{(4)}&\simeq&\left(-1+\frac{2 \sum_i M_i} {r}\right)\left[dt+\frac{2\sum_i{\bm J_i}\cdot ({\bm x}\times d{\bm x})}{r^3} \right]^2+\left(1+\frac{2 \sum_i M_i} {r}\right) \left[ dr^2+r^2 (d\theta^2+\sin^2 \theta d\phi^2) \right],\\
{\bm A}&\simeq& \frac{-2\sum_iQ_i }{ r}dt-2\left(\sum_iP_i \right)\cos\theta d\phi,\\
\phi &\simeq& \frac{2\sqrt{3} \cos2\alpha \sum_i M_i}{r}.
\end{eqnarray}
Thus, the dimensionally reduced spacetime is asymptotically flat.
The Arnowitt-Deser-Misner (ADM) mass, ADM angular momentum, and the total electric and magnetic charges are written respectively as
\begin{eqnarray}
&&M=\sum_{i=1}^NM_i,\quad {\bm J}=\sum_{i=1}^N{\bm J_i},
\end{eqnarray}
\begin{eqnarray}
&&P=\sum_{i=1}^NP_i=2M\cos^3\alpha,\quad Q=\sum_{i=1}^N Q_i=2M \sin^3\alpha,
\end{eqnarray}
where each individual electric and magnetic charge satisfies
\begin{eqnarray}
(P_i,Q_i)=2M_i (\cos^3\alpha,\sin^3\alpha),
\end{eqnarray}
leading to the constraint
\begin{eqnarray}
\left(\frac{P}{2M}\right)^{\frac{2}{3}}+\left(\frac{Q}{2M}\right)^{\frac{2}{3}}=1.\label{eq:ratio}
\end{eqnarray}

\subsection{Regularity}
If curvature singularities exist on or outside the horizons, they must occur at points where the metric or its inverse diverges. 
This occurs only on the surfaces $H_+(x,y,z)=0$ or $H_-(x,y,z)=0$.
Indeed, the Kretschmann scalar behaves as
\begin{eqnarray}
R^{(4)}_{\mu\nu\rho\lambda}R^{(4)\mu\nu\rho\lambda}\sim \frac{1}{(H_+H_-)^3},
\end{eqnarray}
so curvature singularities arise precisely on the surfaces where either $H_+(x,y,z)=0$ or $H_-(x,y,z)=0$. 
We now show that such singularities do not exist on or outside the event horizons at ${\bm x}={\bm x}_i$, provided that $|{\bm J_i}|<|P_iQ_i|$.
To this end, it suffices to verify that $H_\pm>0$ on and outside the horizons, since asymptotically $H_\pm \to 1>0$ as $r\to\infty$.
First, under the assumption $M_i>0$, we have $f>0$, while $g$ can take both signs. 
Introducing $(c_2,s_2):=(\cos2\alpha,\sin2\alpha)$, we rewrite $H_\pm$ as
\begin{eqnarray}
H_\pm   &=& 1+2(1\pm c_2)f+s_2^2(1\pm c_2) f^2+2s_2^2(c_2\pm1)g\notag\\
              &>& s_2^2(1\pm c_2)+2s_2^2(1\pm c_2)f+s_2^2(1\pm c_2) f^2+2s_2^2(c_2\pm1)g\mp s_2^2c_2\notag\\
              &=& s_2^2(1\pm c_2 )[(1+f)^2\pm2g]\mp s_2^2c_2\notag\\
              &\ge& s_2^2(1\pm c_2 )[(1+f)^2- 2|g|]\mp s_2^2c_2\notag\\
              &>& s_2^2(1\pm c_2 )\mp s_2^2c_2\notag\\
              &=& s_2^2>0, \label{eq:Hpm>0}
\end{eqnarray}
where we have used the inequality
\begin{eqnarray}
(1+f)^2-2|g|&=&\left(1+\sum_if_i\right)^2-2\left|\sum_i g_i \right|\notag\\
                  &>&1+\sum_if_i^2-2\sum_i \left|g_i  \right|\notag\\
                  &=&1+\sum_i  \frac{M_i^2}{|{\bm x}-{\bm x}_i|^2}  -\sum_i \frac{M_i^2|{\bm J}_i\cdot ({\bm x}-{\bm x_i})|}{|P_iQ_i||{\bm x}-{\bm x_i}|^3}  \notag\\
                   &\ge&{1+\sum_i  \frac{M_i^2}{|{\bm x}-{\bm x}_i|^2}  -\sum_i \frac{M_i^2|{\bm J_i}|}{|P_iQ_i||{\bm x}-{\bm x_i}|^2}  }\notag\\
                  &=&1+\sum_i M_i^2\left[\frac{1-|{\bm J_i}|/|P_iQ_i|}{|{\bm x}-{\bm x}_i|^2}\right]\notag\\
                  &>&1,
\end{eqnarray}
with
\begin{eqnarray}
f_i:=\frac{M_i}{|{\bm x}-{\bm x_i}|},\quad g_i:=\frac{M_i^2{\bm J}_i\cdot ({\bm x}-{\bm x_i})}{2P_iQ_i|{\bm x}-{\bm x_i}|^3}.
\end{eqnarray}

\subsection{Absence of CTCs}

Here, we demonstrate that no closed timelike curves (CTCs) exist anywhere on or outside the horizons, provided that the inequality $|{\bm J_i}| < |P_i Q_i|$ holds for each $i = 1, \ldots, N$.
The absence of CTCs is equivalent to requiring that the $3\times 3$ matrix
$g^{(4)}_{IJ}$ $(I, J = x, y, z)$ for the 4D metric~(\ref{eq:sol:4Dmetric}) be positive definite everywhere on and outside the horizons. 
By Sylvester's criterion, the matrix $(g_{IJ})$ is positive definite if and only if all its leading principal minors are positive:
\begin{eqnarray}
g^{(4)}_{xx}&=&\frac{H_+H_--(\omega_{x}^0)^2}{\sqrt{H_+H_-}}, \label{eq:ineq1b}\\
{\rm det} 
\begin{pmatrix}
g^{(4)}_{xx} & g^{(4)}_{xy} \\
g^{(4)}_{xy} & g^{(4)}_{yy} \\
\end{pmatrix}
&=&H_+H_--(\omega^0_x)^2-(\omega^0_y)^2>0,  \label{eq:ineq2b}\\
{\rm det} 
\begin{pmatrix}
g^{(4)}_{xx} & g^{(4)}_{xy} & g^{(4)}_{xz}\\
g^{(4)}_{xy} & g^{(4)}_{yy} & g^{(4)}_{yz}\\
g^{(4)}_{xz} & g^{(4)}_{yz} & g^{(4)}_{zz}\\
\end{pmatrix}
&=&\sqrt{H_+H_-}\left[
H_+H_--(\omega^0_x)^2-(\omega^0_y)^2-(\omega^0_z)^2\right]>0. 
\label{eq:ineq3b}
\end{eqnarray}
Under the condition~(\ref{eq:Hpm>0}),  it is straightforward to verify that if the condition~(\ref{eq:ineq3b}) holds, then the conditions, the inequalities~(\ref{eq:ineq1b}) and (\ref{eq:ineq2b}) are automatically satisfied. 
Therefore, it suffices to consider only the inequality~(\ref{eq:ineq3b}).

\medskip
The left-hand side of (\ref{eq:ineq3b})  divided by $\sqrt{H_+H_-}$ can be expressed as
\begin{eqnarray}
H_+H_--(\omega^0_x)^2-(\omega^0_y)^2-(\omega^0_z)^2&=&1+4f+s_2^2(6f^2+4c_2g)+4s_2^4f^3+s_2^6(f^4-4g^2)\notag\\
&&-(\omega^0_x)^2 - (\omega^0_y)^2-(\omega^0_z)^2\\
&>&1+\sum_i[s_2^6(f_i^4-4g_i^2)-(\omega^0_{xi})^2-(\omega^0_{yi})^2-(\omega^0_{zi})^2]\notag\\
&&+2\sum_{i< j}[s_2^6(3f_i^2f_j^2-4g_ig_j)-(\omega^0_{xi})(\omega^0_{xj})-(\omega^0_{yi})(\omega^0_{yj})-(\omega^0_{zi})(\omega^0_{zj})],
\end{eqnarray}
with
\begin{eqnarray}
\omega^0_{xi}=\frac{2[{\bm J_{i}}\times ({\bm x}-{\bm x_i})]_x}{|{\bm x}-{\bm x_i}|^3},\quad 
\omega^0_{yi}=\frac{2[{\bm J_{i}}\times ({\bm x}-{\bm x_i})]_y}{|{\bm x}-{\bm x_i}|^3},\quad 
\omega^0_{zi}=\frac{2[{\bm J_{i}}\times ({\bm x}-{\bm x_i})]_z}{|{\bm x}-{\bm x_i}|^3},\quad 
\end{eqnarray}
where the second and third summations can be shown to be positive under the conditions $|{\bm J_i}|<|P_iQ_i|\ (i=1,\ldots,N)$. Indeed,
\begin{eqnarray}
s_2^6(f_i^4-4g_i^2)-(\omega^0_{xi})^2-(\omega^0_{yi})^2-(\omega^0_{zi})^2&=&s_2^6\left(\frac{M_i^4}{|{\bm x}-{\bm x_i}|^4}-\frac{M_i^4|{\bm J_i}\cdot ({\bm x}-{\bm x_i})|^2}{P_i^2Q_i^2|{\bm x}-{\bm x_i}|^6}\right)
-\frac{4|{\bm J_i}\times ({\bm x}-{\bm x_i})|^2}{|{\bm x}-{\bm x_i}|^6}\notag\\
&=&s_2^6\left(\frac{M_i^4}{|{\bm x}-{\bm x_i}|^4}-\frac{M_i^4|{\bm J_i}\cdot ({\bm x}-{\bm x_i})|^2}{P_i^2Q_i^2|{\bm x}-{\bm x_i}|^6}\right)
-s_2^6\frac{M_i^4|{\bm J_i}\times ({\bm x}-{\bm x_i})|^2}{P_i^2Q_i^2|{\bm x}-{\bm x_i}|^6}\notag\\
&=&s_2^6M_i^4\left(\frac{1}{|{\bm x}-{\bm x_i}|^4}-\frac{|{\bm J_i}\cdot ({\bm x}-{\bm x_i})|^2+|{\bm J_i}\times ({\bm x}-{\bm x_i})|^2}{P_i^2Q_i^2|{\bm x}-{\bm x_i}|^6}\right)\notag\\
&=&s_2^6M_i^4\left(\frac{1}{|{\bm x}-{\bm x_i}|^4}-\frac{|{\bm J_i}|^2|{\bm x}-{\bm x_i}|^2}{P_i^2Q_i^2|{\bm x}-{\bm x_i}|^6}\right)\notag\\
&=&s_2^6\frac{M_i^4(1-|{\bm J_i}|^2/|P_iQ_i|^2)}{|{\bm x}-{\bm x_i}|^4}\notag\\
&>&0,
\end{eqnarray}
and
\begin{eqnarray}
&&s_2^6(3f_i^2f_j^2-4g_ig_j)-(\omega^0_{xi})(\omega^0_{xj})-(\omega^0_{yi})(\omega^0_{yj})-(\omega^0_{zi})(\omega^0_{zj})\notag\\
&&=s_2^6\left(\frac{3M_i^2M_j^2}{|{\bm x}-{\bm x_i}|^2|{\bm x}-{\bm x_j}|^2}
-\frac{M_i^2M_j^2\{{\bm J_i}\cdot ({\bm x}-{\bm x_i})\} \{{\bm J_j}\cdot ({\bm x}-{\bm x_j})\}\ }{P_iP_jQ_iQ_j|{\bm x}-{\bm x_i}|^3|{\bm x}-{\bm x_j}|^3}\right)
-\frac{4\{{\bm J_i}\times ({\bm x}-{\bm x_i})\}\cdot \{{\bm J_j}\times ({\bm x}-{\bm x_j})\}  }{|{\bm x}-{\bm x_i}|^3|{\bm x}-{\bm x_j}|^3}\notag\\
&&=s_2^6M_i^2M_j^2\left(\frac{3}{|{\bm x}-{\bm x_i}|^2|{\bm x}-{\bm x_j}|^2}
-\frac{\{{\bm J_i}\cdot ({\bm x}-{\bm x_i})\} \{{\bm J_j}\cdot ({\bm x}-{\bm x_j})\}+\{{\bm J_i}\times ({\bm x}-{\bm x_i})\}\cdot \{{\bm J_j}\times ({\bm x}-{\bm x_j})\} }{P_iP_jQ_iQ_j|{\bm x}-{\bm x_i}|^3|{\bm x}-{\bm x_j}|^3}\right)\notag\\
&&\ge s_2^6M_i^2M_j^2\left(\frac{3}{|{\bm x}-{\bm x_i}|^2|{\bm x}-{\bm x_j}|^2}
-\frac{|\{{\bm J_i}\cdot ({\bm x}-{\bm x_i})\} \{{\bm J_j}\cdot ({\bm x}-{\bm x_j})\}+\{{\bm J_i}\times ({\bm x}-{\bm x_i})\}\cdot \{{\bm J_j}\times ({\bm x}-{\bm x_j})\} | }{|P_iP_jQ_iQ_j||{\bm x}-{\bm x_i}|^3|{\bm x}-{\bm x_j}|^3}\right)\notag\\
&& \ge s_2^6M_i^2M_j^2\left(\frac{3}{|{\bm x}-{\bm x_i}|^2|{\bm x}-{\bm x_j}|^2}
-\frac{|\{{\bm J_i}\cdot ({\bm x}-{\bm x_i})||{\bm J_j}\cdot ({\bm x}-{\bm x_j})|+|{\bm J_i}\times ({\bm x}-{\bm x_i})||{\bm J_j}\times ({\bm x}-{\bm x_j})| }{|P_iP_jQ_iQ_j||{\bm x}-{\bm x_i}|^3|{\bm x}-{\bm x_j}|^3}\right)\notag\\
&&= s_2^6M_i^2M_j^2\left(\frac{3}{|{\bm x}-{\bm x_i}|^2|{\bm x}-{\bm x_j}|^2}
-\frac{|{\bm J_i}||{\bm J_j}| |{\bm x}-{\bm x_i}||{\bm x}-{\bm x_j}|(|\cos\theta_i\cos\theta_j|+|\sin\theta_i\sin\theta_j|) }{|P_iP_jQ_iQ_j||{\bm x}-{\bm x_i}|^3|{\bm x}-{\bm x_j}|^3}\right)\notag\\
&&\ge s_2^6M_i^2M_j^2\left(\frac{3}{|{\bm x}-{\bm x_i}|^2|{\bm x}-{\bm x_j}|^2}-\frac{|{\bm J_i}||{\bm J_j}| }{|P_iP_jQ_iQ_j||{\bm x}-{\bm x_i}|^2|{\bm x}-{\bm x_j}|^2}\right)\notag\\
&&\ge s_2^6\frac{3M_i^2M_j^2(1-|{\bm J_i}|/|P_iQ_i||{\bm J_j}|/|P_jQ_j|)}{|{\bm x}-{\bm x_i}|^2|{\bm x}-{\bm x_j}|^2}\notag\\
&&>0.
\end{eqnarray} 
Here, $\theta_i$ and $\theta_j$ denote the angles between ${\bm J_i}$ and ${\bm x}-{\bm x_i}$, 
and between ${\bm J_j}$ and ${\bm x}-{\bm x_j}$, respectively. 
In deriving the third inequality from the end, we have used the Cauchy--Schwarz inequality
\begin{eqnarray}
|\cos\theta_i\cos\theta_j|+|\sin\theta_i\sin\theta_j|
= {\bm a}\cdot {\bm b} \le |{\bm a}||{\bm b}| = 1,
\end{eqnarray}
where ${\bm a}=(|\cos\theta_i|,|\sin\theta_i|)$ and ${\bm b}=(|\cos\theta_j|,|\sin\theta_j|)$.

\medskip
The positivity of the above expression implies that the inequality~(\ref{eq:ineq3b}) holds under the conditions 
$|{\bm J_i}|<|P_iQ_i|\ (i=1,\cdots,N)$.
Therefore, no closed timelike curves (CTCs) exist on or outside the horizons.

\subsection{Two black holes}\label{sec:2BH}
Finally, we consider the case of two rotating black holes with non-aligned angular momenta.
For the case with aligned angular momenta, the symmetry of the spacetime is enhanced when two black holes are  placed on the $z$-axis, since the geometry then admits an additional rotational Killing vector generated by the rotation about the axis.
However, for the non-aligned angular momenta, this is not the case. 
Therefore, the spacetime has only a single Killing vector, a timelike Killing vector, does not have the spacelike Killing vector.

\medskip
We now present the explicit form of the solution in the case of two black holes. 
Without loss of generality, we take their positions to be ${\bm x}_1=(0,0,-a)$ 
and ${\bm x}_2=(0,0,a)$. 
Introducing cylindrical coordinates $(\rho,\phi,z)$ defined by $(x,y,z)=(\rho\cos\phi,\rho\sin\phi,z)$, we obtain
\begin{eqnarray}
ds^2_{(4)}&=&-\frac{1}{\sqrt{H_+H_-}}(dt+{\bm \omega^0})^2+\sqrt{H_+H_-}(d\rho^2+\rho^2d\phi^2+dz^2),
\label{eq:sol_2BH}
\end{eqnarray}
with (\ref{eq:Hpma}) and 
{\small 
\begin{eqnarray}
f&=&\frac{M_1}{\sqrt{\rho^2+(z+a)^2}}+\frac{M_2}{\sqrt{\rho^2+(z-a)^2}},\label{eq:f2}\\
g&=&\frac{M_1^2}{2P_1Q_1}\frac{ \rho(J_{x,1}\cos\phi+J_{y,1}\sin\phi)+J_{z,1}(z+a)  }{ \sqrt{\rho^2+(z+a)^2}^3}
+\frac{M_2^2}{2P_2Q_2}\frac{ \rho(J_{x,2}\cos\phi+J_{y,2}\sin\phi)+J_{z,2}(z-a)  }{ \sqrt{\rho^2+(z-a)^2}^3},\label{eq:g2}\\
{\bm \omega^0}&=&2\frac{[J_{y,1} (z+a)-J_{z,1}y ] dx+[J_{z,1} x-J_{x,1}(z+a) ] dy+[J_{x,1} y-J_{y,1}x ] dz}{\sqrt{\rho^2+(z+a)^2}^3}\notag\\
&&+2\frac{[J_{y,2} (z-a)-J_{z,2}y ] dx+[J_{z,2} x-J_{x,2}(z-a) ] dy+[J_{x,2} y-J_{y,2}x ] dz}{\sqrt{\rho^2+(z-a)^2}^3}\\
&=&\small 2\frac{[(J_{y,1} \cos\phi -J_{x,1} \sin\phi](z+a)d\rho 
+\rho[    J_{z,1}\rho -(J_{x,1}\cos\phi+J_{y,1} \sin\phi)(z+a)                         ]d\phi
+\rho(J_{x,1} \sin\phi-J_{y,1}\cos\phi )dz}{\sqrt{\rho^2+(z+a)^2}^3}\notag\\
&+&2\frac{[(J_{y,2} \cos\phi -J_{x,2} \sin\phi](z-a)d\rho 
+\rho[    J_{z,2}\rho -(J_{x,2}\cos\phi+J_{y,2} \sin\phi)(z-a)                          ]d\phi
+\rho(J_{x,2} \sin\phi-J_{y,2}\cos\phi )dz}{\sqrt{\rho^2+(z-a)^2}^3},\notag\\
\label{eq:omega02}
\end{eqnarray}
}
where $\partial_\phi$ does not correspond to a rotational Killing vector about the $z$-axis.

 \medskip
 In the case of non-aligned angular momenta, the spacetime does not admit a rotational Killing vector, and therefore the Weyl--Papapetrou coordinates cannot be employed.
In contrast, for aligned angular momenta, ${\bm J}_1=(0,0,J_{z,1})$ and ${\bm J}_2=(0,0,J_{z,2})$, the spacetime becomes axisymmetric and admits the rotational Killing vector $\partial_\phi$. In this case, the coordinates $(\rho,z)$ can be identified with the Weyl--Papapetrou coordinates. Consequently, one can analyze the properties of the solution using the rod structure~\cite{Harmark:2004rm}. 
The $z$-axis can be divided into three segments by the two black holes located at $z=\pm a$ as follows:
\begin{itemize}
\item[(i)] rotational axis: $I_1=\{(\rho,z)\mid \rho=0,-\infty<z<-a\}$, with the rod vector $(0,1)$, 
\item[(ii)] rotational axis: between the horizons $I_2=\{(\rho,z)\mid \rho=0,-a<z<a\}$,  with the rod vector $(0,1)$, 
\item[(iii)] rotational axis: $I_3=\{(\rho,z)\mid \rho=0,a<z<\infty\}$, with the rod vector $(0,1)$,
\end{itemize}
In this case, the one-form in (\ref{eq:omega02}) can be written in the simple form ${\bm \omega_0}=\omega_\phi, d\phi$, and $\omega_\phi$ vanishes on $I_1$, $I_2$, and $I_3$. This implies that no Dirac--Misner string singularities arise.
Furthermore, following the standard method (see~\cite{Harmark:2004rm}, for instance) , one finds that on $I_1$, $I_2$, and $I_3$.
\begin{eqnarray}
\lim_{\rho \to 0} \sqrt{\frac{\rho^2 g_{\rho\rho}}{g_{\phi\phi}}} =1,
\end{eqnarray}
which shows that no conical singularities are present on $I_1$, $I_2$, and $I_3$.

\medskip
Thus, we can establish the absence of conical singularities in the case of aligned angular momenta. 
However, the above analysis cannot be directly extended to the case of non-aligned angular momenta.
Since the near-horizon geometries around $z = \pm a$ reduce to that of a singly rotating extremal Rasheed--Larsen black hole in the slow-rotation limit, conical singularities are absent, at least, on the horizons at $(\rho,z) =(0, \pm a)$. This observation suggests that they may also be absent both on and off the $z$-axis.
At present, we do not know how to demonstrate this rigorously, and we leave this issue for future work.

\section{Summary and Discussion}\label{sec:summary}

In this paper, we have constructed an exact solution describing multi-rotating black holes with non-aligned angular momenta in 5D Kaluza--Klein theory. 
After dimensional reduction along the fifth spatial direction, the solution corresponds to a class of multi-centered rotating dyonic black holes in 4D Einstein--Maxwell--dilaton theory.
The present solution generalizes previously constructed multi-black hole solution with aligned or anti-aligned angular momenta to the case of fully non-aligned spins. 
It is characterized by two harmonic functions on three-dimensional Euclidean space, which determine both the gravitational and electromagnetic fields. 
In contrast to earlier constructions, the angular momenta of individual black holes are allowed to take arbitrary orientations, leading to a richer class of stationary configurations.
We have shown that the resulting spacetimes are asymptotically flat and free from curvature singularities and closed timelike curves on and outside the horizons, provided that the inequality $|{\bm J_i}| < |P_i Q_i|$ is satisfied for each center. The near-horizon geometry is regular, and each black hole possesses an extremal horizon.
The asymptotic structure of the solution has been analyzed, and the total ADM mass, angular momentum, and electric and magnetic charges are obtained as sums of the corresponding quantities of each constituent. 
Finally, we have examined the case of two black holes with non-aligned angular momenta. In this configuration, the spacetime no longer admits a rotational Killing vector, and only a timelike Killing vector remains. 
This feature highlights a qualitative difference from the aligned case, where axial symmetry is preserved.
Our results provide a new class of exact multi-black-hole solutions in 4D or higher-dimensional gravity and offer a useful framework for exploring the interplay between rotation, charge, and regularity in non-supersymmetric configurations.

\medskip
In this work, we have discussed the possibility that conical singularities may exist along the axis between the two horizons in the two-black-hole case.
In axisymmetric configurations, such as the Majumdar--Papapetrou dihole~\cite{Majumdar:1947eu,Papapetrou} and the dihole with aligned or anti-aligned angular momenta~\cite{Teo:2023wfd,Tomizawa:2025tvb}, where the angular momenta are directed along the $z$-axis, the spacetime admits rotational symmetry, and one can demonstrate the absence of conical singularities.
However, the present non-aligned configuration does not possess such symmetry.
Therefore, the standard method (see~\cite{Harmark:2004rm}, for instance) applicable to axisymmetric systems in Weyl-Papapetrou coordinates, where the condition to cure the conical singularities is given by $ \frac{\Delta \phi}{2\pi} =\lim_{\rho \to 0} \sqrt{\frac{\rho^2 g_{\rho\rho}}{g_{\phi\phi}}} $  ($\Delta \phi$ : periodicity), cannot be employed here.
From the analysis of the near-horizon geometry, we find that no conical singularities are present, at least on the horizons. This suggests that they are also absent throughout the spacetime. A more rigorous demonstration of this statement is left for future work.
For the present solution, it is also of physical interest to identify the forces that balance each other in the absence of conical singularities. In this case, the situation is more involved. In addition to gravitational attraction and electrostatic repulsion, dipole-dipole and spin-spin interactions are present. The latter was first discussed by Wald~\cite{Wald:1972sz} using Papapetrou's equation for a spinning particle, where it was shown that, in an appropriate limit, this force takes the same form as the usual dipole-dipole interaction in magnetostatics. Moreover, interactions mediated by scalar fields, namely the dilaton, also affect the force balance.
A detailed analysis of these effects is left for future work.

\medskip
Several interesting directions remain for future investigation.
First, it would be important to extend the present construction to fully non-extremal configurations. 
In the current solution, each black hole is extremal, and it remains an open problem whether multi-centered solutions with non-aligned angular momenta can be realized beyond the extremal limit without introducing singularities or pathologies.
Second, the stability of the obtained solutions should be analyzed. 
In particular, it would be valuable to study linear perturbations and determine whether the absence of closed timelike curves and other regularity conditions is preserved under dynamical fluctuations. 
Third, it would be interesting to investigate whether similar constructions can be achieved in pure 5D vacuum gravity without relying on dimensional reduction. 
This is closely related to the long-standing problem of constructing regular multi-black-hole solutions 
in asymptotically flat vacuum spacetimes~\cite{Tan:2003jz,Herdeiro:2008en}.
Another important direction is to explore the role of hidden symmetries and integrable structures. 
Since the present solution is based on harmonic functions, it may be possible to reformulate the 
construction within the framework of nonlinear sigma models or solution-generating techniques such as 
inverse scattering or the Breitenlohner--Maison formalism.
Furthermore, it would be worthwhile to study the geometric and topological properties of the solution in more detail, particularly the structure of the horizons and their possible generalizations to non-asymptotically flat asymptotics. 
Finally, understanding the physical implications of non-aligned angular momenta, especially their effects on interactions between black holes, may provide new insights into multi-black-hole dynamics in 4D or higher-dimensional gravity.
We hope that the present work provides a useful starting point for these investigations and contributes to the broader understanding of multi-centered configurations in 4D or higher-dimensional gravitational theories.

\acknowledgments
RS was supported by JSPS KAKENHI Grant Number JP18K13541. 
The work of JS was supported by the JSPS Grant-in-Aid for Transformative Research Areas (A) “Extreme Universe” No. 21H05190.




\end{document}